# Circularly Polarized Lasing from a Microcavity Filled with Achiral Single-Crystalline Microribbons

Qian Liang,[a] Xuekai Ma,[b] Teng Long,[a] Jiannian Yao,[c] Qing Liao,[a,*] Hongbing Fu[a,*]

[a]  Q Liang, T Long, Prof Q Liao, Prof H. B.Fu
    Beijing Key Laboratory for Optical Materials and Photonic Devices, Department of Chemistry
    Capital Normal University, Beijing 100048, China
    E-mail: liaoqing@cnu.edu.cn; hbfu@cnu.edu.cn
[b]  Dr. X. Ma
    Department of Physics and Center for Optoelectronics and Photonics Paderborn (CeOPP)
    Universität Paderborn, Warburger Strasse 100, 33098 Paderborn, Germany
[c]  Prof. J. N. Yao
    Beijing National Laboratory for Molecular Sciences (BNLMS)
    Institute of chemistry, Chinese Academy of Sciences, Beijing 100190, China

**Abstract:** Organic circularly polarized (CP) lasers have received increasing attention due to their future photoelectric applications. Here, we demonstrate a CP laser from a pure organic crystal-filled microcavity without any chiral molecules or chiral structures. Benefited from the giant anisotropy and excellent laser gain of organic crystals, optical Rashba-Dresselhaus spin-orbit coupling effect can be induced and is conductive to the CP laser in such microcavities. The maximum dissymmetry factor of the CP lasing with opposite helicities reached, is as high as 1.2. Our strategy may provide a new idea for the design of CP lasers towards future 3D laser displays, information storage and other fields.

## Introduction

Organic circularly polarized (CP) light with the optical-rotatory and angle-independence characteristics, have received increasing attention for their potential applications, such as in nanophotonics,[1] quantum optics,[2] biophysics,[3] and three-dimensional (3D) display technologies.[4] Nevertheless, incoherent broad-band emission of traditional CP light usually brings about the poor spectral resolution and color reproducibility. Organic CP laser, which combine the advantages of organic lasers and CP light emission, exhibits a great potential in areas as information encryption and 3D laser display.[5] The key point in designing efficient organic CP laser is to develop active CP-emissive materials with high optical gain property. Although organic solid-state laser has made significant progress benefiting from the merits of abundant molecular species and high solid photoluminescence (PL) quantum yield, realizing efficient organic CP lasers still remains a critical challenge because that the common organic laser media exhibit the weak CP-emissive feature. One typical strategy to CP laser is the conversion from linearly polarized laser to CP one by use of optical retardation, in which the large power loss induced by the devices is an obstacle to achieving their applications.[6] Another method is the direct generation of CP laser from the chiral compound. However, limited by the difficulties in the time-consuming molecular synthesis and elaborate assembly processes, the chiral materials with high CP-emissive activity[7] are difficult to take into account the efficient stimulated emission character, which narrows the range of organic materials available for CP lasers. Therefore, it is urgent to call for novel organic materials and circular polarization generation mechanism and strategies for developing advanced organic CP lasers.

An alternative approach is the direct manipulation of the polarization degree of freedom or pseudospin, akin to the spin property in electronic systems of photons by means of optical spin-orbit coupling (SOC) in photonic systems.[8] Optical SOC has been demonstrated to enable the engineering of photonic pseudospin states which has led to rapid advances in topological photonics.[9] Recently, a persistent photonic spin texture (i.e., the left- and right-handed circular polarization splitting), a characteristic feature of Rashba-Dresselhaus (RD) SOC with equal Rashba and Dresselhaus coupling strength, was observed in a liquid-crystal-filled optical cavity.[10] By inducing an organic laser dye or perovskite microplates into liquid-crystal cavities, two groups have observed independently the phenomenon of CP lasing.[11] Organic single-crystals (OSCs) are of current interest in laser applications because of their chemically tunable optoelectronic properties, self-assembly amenability to low-cost fabrication, and largely stimulated emission cross-section.[12] Particularly, OSCs possess the intrinsic feature of long-range ordered molecular stacking arrangement, which leads to crystals to not only free from traps and grain boundaries but also to a distinct anisotropy of the refractive-index distribution. In fact, this anisotropic refractive index has been used to engineer synthetic RD Hamiltonians in an achiral perylene crystal-filled microcavity and to realize the scheme of CP dispersion via the passive reflection mode.[13] Therefore, integrating the advantages of excellent lasing characteristics and engineering of pseudospin-polarized textures, OSC-filled microcavities may open a new avenue for active CP lasers without the need for chiral organic molecules or chiral structures. So far, to the best of our knowledge, organic CP lasers assisted by RD SOC in OSC-filled microcavities have not been explored yet.

In the present paper we demonstrate CP lasing emission with a high dissymmetry factor (*g*) of 1.2 from an organic microcavity filled by an achiral single-crystalline microribbon of 1,4-bis((E)-2,4-dimethylstyryl)-2,5-dimethylbenzene (6M-DSB, Figure 1a). The circular polarization splitting of the optical modes induced by photonic RD SOC is observed experimentally in the momentum space. Combining the excellent lasing gain feature of 6M-DSB single-crystalline microribbons with the artificial RD SOC Hamiltonian, we realize a CP lasing emission from the 6M-DSB OSC-filled cavity without any chiral molecules or chiral structures. These results provide an alternative molecular design strategy for the development of organic CP lasers towards novel 3D laser display and information storage.

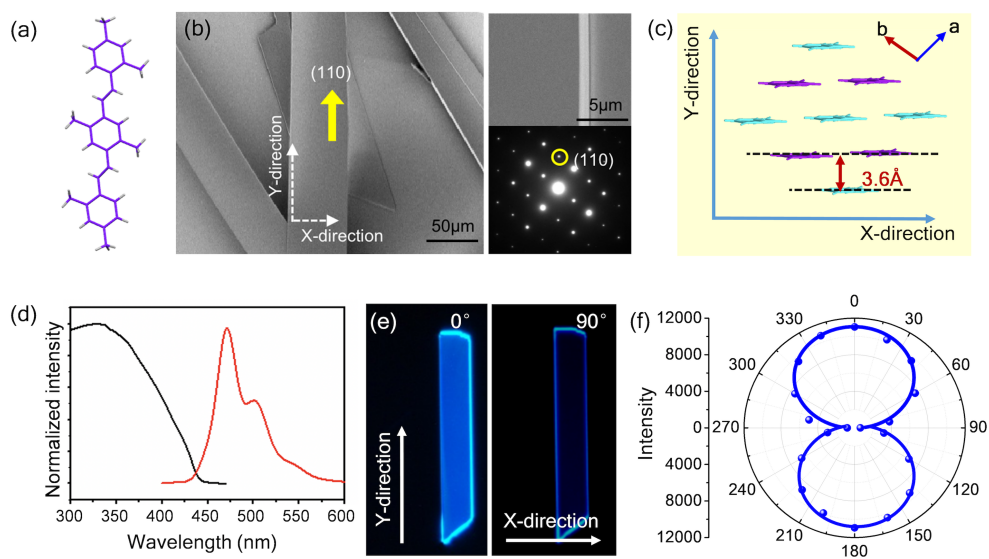

**Figure 1.** (a) The molecular structure of 6M-DSB. (b) SEM image of a typical 6M-DSB OSC microbelt. Insets: The magnified SEM image (upper) and SAED image (bottom). (c) Brickwork molecular packing arrangement within the (001) crystal plane, viewed perpendicular to the microribbon top-facet. (d) Normalized steady state absorption (black) and PL (red) spectra of 6M-DSB microribbons. Excitation-light-polarization dependent fluorescence microscope images (e) and the corresponding PL emission intensities (f) of a single microribbon, recorded by rotating the angle between the excitation-laser polarization and Y-direction.

## Results and Discussion

The microstructures of 6M-DSB are obtained by a facile micro-spacing in-air sublimation method[14] and the corresponding apparatus is schematized in Scheme S1 (see details in Supporting Information). Figure 1b shows a scanning electron microscope (SEM) image of typical products deposited on the silicon wafer. It can be seen that as-prepared 6M-DSB samples exhibit the ribbon-like structure with smooth outer surfaces and sharp edges. The edge length is found to range from tens to hundreds of microns and the width is about 20-80 µm. The thickness is around ~ 1 µm with a roughness less than 1 nm according to atomic force microscopy (AFM) measurements (Figure S1), which is conductive to fabricate optical microcavities. The selected-area electron diffraction (SAED) and X-ray diffraction (XRD) measurements were carried out to understand the packing arrangement of 6M-DSB molecules (CCDC 1043236) in OSCs. The sharp spots in SAED image (bottom inset of Figure 1b) and XRD pattern (Figure S2) reveal that these microribbons are single-crystalline structures. In these OSCs, 6M-DSB molecules adopt a lamellar structure with the crystal (001) plane parallel to the substrate.[14] Within the (001) plane, nearly planar 6M-DSB molecules stand on the substrate and stack in a brickwork arrangement with the nearest π-π distance of ca. 3.6 Å (Figure 1c). This brickwork arrangement brings about a significantly anisotropic molecular packing density along and parallel to the π-π interaction (defined as Y- and X-direction, respectively), thus leading to a strong anisotropy of refractive-index distribution in the 6M-DSB OSCs. We associate the linearly polarized modes along X- and Y-direction as X- and Y-polarization, respectively.

Figure 1d shows the normalized steady-state absorption and PL spectra of 6M-DSB OSCs. Their absorption spectrum exhibits a wide absorption with a maximum at 330 nm (black line), while the PL spectrum is dominated by 0-1 transition at 470 nm, which are in good agreement with our previous report.[15] In order to perform the anisotropic properties of 6M-DSB OSCs, we investigate the characteristics of linear polarization-dependent PL from an individual microribbon by the excitation-light-polarization-dependent spectroscopy (Figure S3). A half wave plate was placed between the sample and light source to modify the polarization of the excitation light. When the polarization of the excitation light is rotated to be parallel (i.e., θ = 0°) or vertical (i.e., θ = 90°) to the Y-direction, respectively, maximum and minimum PL emission are detected (Figure 1e). We recorded PL emission intensity as a function of the different rotation angles and shown in Figure 1f. The obvious variation of the PL intensities allows us to calculate the polarization degree, defined by ($I_{max}$ - $I_{min}$) / ($I_{max}$ + $I_{min}$), as high as 0.89, where $I_{max}$ and $I_{min}$ correspond respectively to the intensities parallel and vertical to Y-direction. This also strongly supports the fact of significant anisotropy of our 6M-DSB OSCs.

We further fabricated the 6M-DSB OSC-filled microcavity and its schematic figure is shown in Figure 2a. The cavity is composed of two silver films with the thickness of 100 nm (reflectivity ≥ 98%) and 35 nm (reflectivity ≥ 50%) separated by a certain distance. This space is filled with a 6M-DSB single-crystalline microribbon. In order to prevent fluorescence quenching caused by direct contact of the 6M-DSB microribbon with the silver reflectors, 20 nm-thick SiO$_2$ thin-films are introduced as space-layers between 6M-DSB and silver films.[16] Due to the large birefringence of 6M-DSB OSCs, for a cavity mode with a parity number $l$, the two orthogonally linearly polarized modes (X$_l$ and Y$_l$) produce an energy splitting at k = 0 in the momentum space (Figure 2b). This energy splitting can be described by the parameter $\beta_0$, which splits the linearly polarized modes with different parity, X$_l$ and Y$_{l-1}$, at k = 0. Note that $\beta_0$ = 0 means that the modes of X$_l$ and Y$_{l-1}$ are resonant at k=0. When X$_l$ and Y$_{l-1}$ modes are resonant, RD SOC appears (Figure 2c).[10c, 13] In this case, the eigenmodes of orthogonally linearly polarized modes split into CP modes along with the change of the dispersion relation because of the RD SOC with equal Rashba and Dresselhaus coupling strength. In the circular polarization basis, such systematic effective Hamiltonian can be described by a 2×2 matrix with the form: [10c,13]

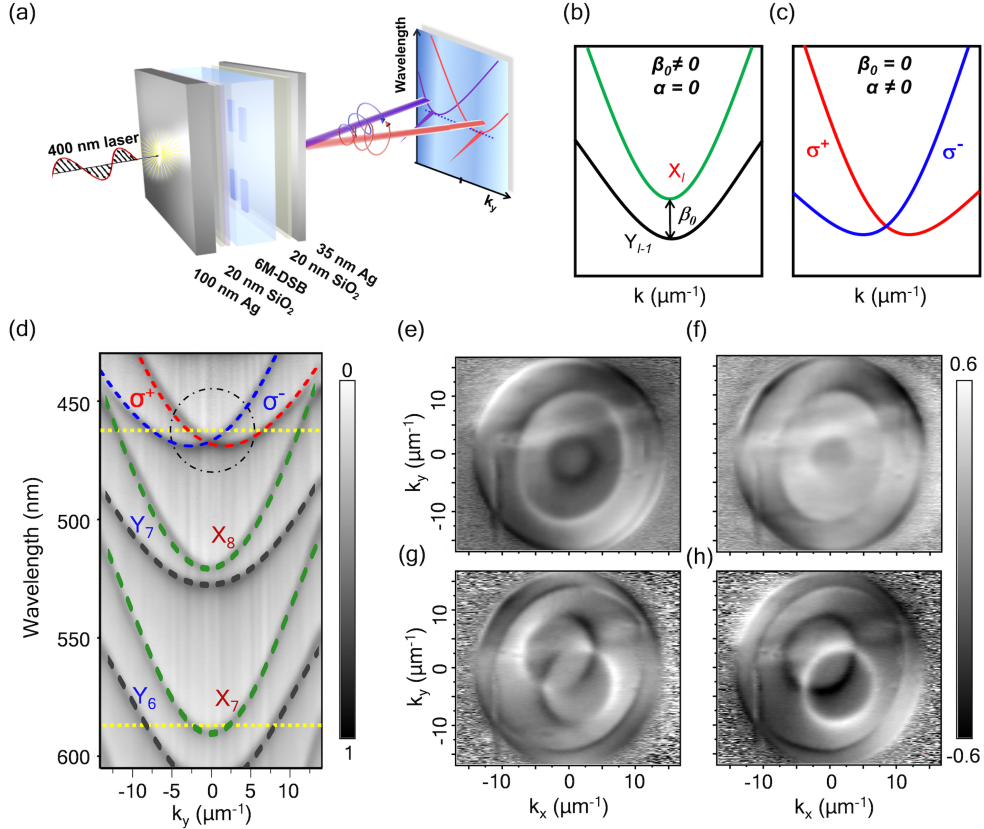

**Figure 2.** (a) Scheme of the OSC-filled microcavity and the lasing process. Upon excitation with a linearly polarized 400-nm laser, lasing emission occurs at the lowest energy of the dispersion relation. (b) Schematics of the dispersion of two orthogonally linearly polarized modes in an anisotropic microcavity. (c) Schematics of the dispersion when RD SOC emerges as two orthogonally linearly polarized modes with opposite parity are resonant. (d) ARR of the microcavity with organic layer thickness of 1020 nm. The green and black dashed curves represent the simulated X and Y cavity modes, respectively. The red and blue lines are the numerical results, representing the circular polarizations, calculated by using Eq. (1) with parameters $E_0$ = 2.655 eV (467 nm), m = $2.3 \times 10^{-5}$ $m_e$ ($m_e$ is the free electron mass), $\beta_0$ = 0, $\beta_1$ = 0.6 meV, and $\alpha$ = 33 eV Å. (e-h) Cross-section maps of the 2D tomography at 465 nm and 585 nm in momentum space, corresponding to the yellow dashed lines in (d), respectively, with $S_1$ components of the Stokes vector at 585 nm (e) and 465 nm (g) and $S_3$ components of the Stokes vector at 585 nm (f) and 465 nm (h).

$$H(\boldsymbol{k}) = \begin{pmatrix} E_0 + \frac{\hbar^2}{2m}\boldsymbol{k}^2 - 2\alpha k_y & \beta_0 + \beta_1 \boldsymbol{k}^2 e^{2i\varphi} \\ \beta_0 + \beta_1 \boldsymbol{k}^2 e^{-2i\varphi} & E_0 + \frac{\hbar^2}{2m}\boldsymbol{k}^2 + 2\alpha k_y \end{pmatrix}. \quad (1)$$

Here, $E_0$ is the energy of the ground state, $m$ is the effective mass of cavity photons, $\boldsymbol{k} = (k_x, k_y)$ is the in-plane wave vector and $\varphi$ ($\varphi \in [0, 2\pi]$) is the in-plane polar angle. $\beta_1$ is the strength of the intrinsic transverse-electric-transverse-magnetic (TE-TM) splitting of the cavity modes,[17] and $\alpha$ is the RD coupling strength.[10a, 10c,13]

In order to observe experimentally the RD spin texture, the unpolarized angle-resolved reflectivity (ARR) of the OSC-filled microcavity with the thickness of 1020 nm were performed by using a home-made micro-scale ARR measurement setup (Scheme S2 and S3) at room temperature. Figure 2d shows the reflectivity as a function of wavelength and in-plane wave vector ($k_y$), in which two sets of modes with distinctive curvatures can be seen (also see Figure S4). We have calculated X and Y modes by using the two-dimensional (2D) cavity photon dispersion relations,[13] which agrees well with experimental results (see the green and black dashed lines in Figure 2d and see also Figure S5). The simulated refractive indices of the two cavity modes (that is, 2.10 for X-polarized modes and 2.50 for Y-polarized ones) also support that the 6M-DSB OSCs have giant anisotropy (Figure S5). Furthermore, the polarization-resolved ARR (energy vs. in-plan wave vector) experiments have been carried out for the six different polarization components of light (horizontal-vertical, diagonal-anti-diagonal, and left and right circular), which allows a full determination of the polarization characteristics (three Stokes vector components) of the cavity modes.[18] Thus, the modes with larger and smaller curvatures of 6M-DSB OSCs are attributed to X- and Y-polarizations, corresponding to X- and Y-direction, respectively.

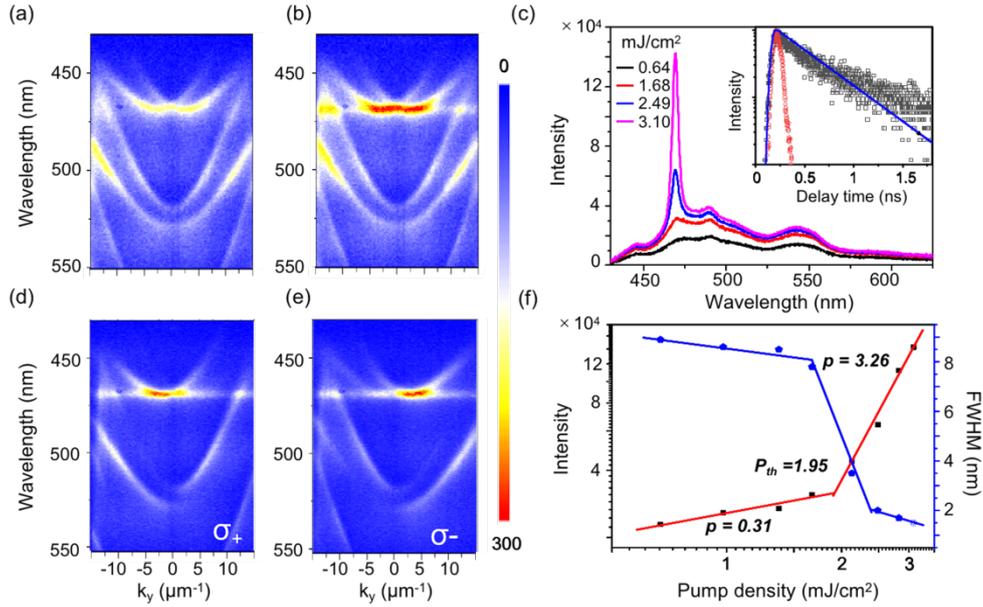

**Figure 3.** (a) Angle-resolved PL spectra below (a) and above the threshold (b). (c) PL emission intensity as a function of wavelength at different pump densities. Inset: PL decay profiles over time for below and above the threshold. Angle-resolved polarized emission spectra of σ+ (d) and σ− (e) components. (f) Emission intensity (red) and FWHM (blue) as a function of the pump density.

In the vicinity of 470 nm, the RD SOC emerges in the OSC-filled cavity and the dispersion curves are well fitted by the numerical results calculated by using Eq. (1). We have extracted the Stokes vector components to analyze the polarization property of the emitted light. The $S_1$ components of the Stokes vector of $X_8$ and $Y_7$ branches present strongly linearly polarized PL emission as shown in the 2D wavevector map of the tomography (Figure 2e), while their corresponding $S_3$ components are relatively weaker (Figure 2f), which indicates that there is no RD SOC effect. As X- and Y-polarized modes approach at $k_y = 0$, such as $X_9$ and $Y_8$ shown in Figure S5, the nearly resonant two modes trigger the RD SOC, leading to a clear splitting of the paraboloids along $k_y$ direction (σ+ and σ- in Figure 2d). The $S_3$ components present two separate circles with opposite signs, that is, opposite circular polarizations (Figure 2g). Meanwhile, the signals in the $S_1$ components become very weak (Figure 2h). These further demonstrate the occurrence of the RD SOC effect. Thus, the circular polarization slitting of the modes at near 470 nm provides a persistent spin helix for PL emission. From this it is expected that CP lasing can be realized in the RD-SOC-assisted scheme by introducing organic materials with sufficient amount of gain.

Figure 3 shows the corresponding lasing behavior of the 6M-DSB OSC-filled microcavity for excitation with a 400-nm femtosecond laser pulse from a Ti:sapphire regenerative amplifier at room temperature. The angle-resolved PL (ARPL) spectra obtained below the lasing threshold (Figure 3a) reveal the dispersion relation of PL emission close to the RD regime, which exactly coincides with the dispersion curves from the ARR spectrum (Figure 2d). It indicates that the PL emission of the OSC-filled cavity should inherit the persistent spin-helix characteristics shown in its absorption spectrum. Figure 3b presents the measurement above the lasing threshold (about 2.1 mJ/cm$^2$). At the bottom of the two spin-polarized valleys ($k = \pm 2.6$ μm$^{-1}$) near 470 nm, enhancement and both spectral and angular narrowing of PL emission take place. Figure 3c shows emission spectra under different pumping powers. A series of broad spontaneous emission peaks are observed under lower pumping density, while a stronger lasing emission develops around 470 nm and finally becomes dominant as the pumping density increases. The emission intensity as a function of pump power present a typical lasing threshold curve (Figure 3f). The clear lasing threshold at $P_{th} = 1.95$ mJ/cm$^2$ is identified by the intersection between the sublinear and superlinear regions. The intensity dependence of these two regions is separately fitted to a power law $x^p$ with $p = 0.31\pm0.02$ and $3.26\pm0.06$, respectively. The FWHM is drastically narrowed from 8.5 nm below the threshold to 1.5 nm above the threshold. Meanwhile, the decay lifetime constant sharply collapses from $\tau = 0.41\pm0.01$ ns below the threshold to $\tau < 15$ ps above the threshold (inset of Figure 3c) based on time-resolved PL measurements using a streak camera. The superlinear increase of the PL intensity, the decreased linewidth of the modes, and the collapse of the lifetime confirm the occurrence of lasing from the bottom of two valleys in the dispersions (Figure 2a). Because both valleys are oppositely circularly polarized, the coherent light emitted from the two valleys also inherits these polarization properties. This effect is evidenced by the polarization-dependent emission measurements. Figure 3d and 3e present strong left- (σ+) and right-hand (σ-) circularly polarized lasing emission near 470 nm in the vicinity of $k_y = \pm 2.5$ μm$^{-1}$, respectively. This strongly testifies that the RD SOC leads to the CP lasing emission due to the circular polarization splitting.

The cavity mode degeneracy is a wavelength-dependent feature, so that the RD SOC can appear at different wavelengths as the cavity thickness varies. Therefore, the polarization degree of lasing emission strongly depends on the relative wavelength difference between the lasing emission point (470 nm) and the crossing point where the RD SOC occurs. Figure 4a and 4c show the typical ARR spectra with different cavity lengths (also see Figure S6). Clearly, the characteristic crossing point of the RD SOC dispersion can be modified in a large spectral range by changing the cavity length. We define the parameter $\Delta\delta$ as the wavelength difference between the lasing emission and the crossing point. Figure 4b and 4d show the corresponding ARPL spectra above their lasing thresholds. All the 6M-DSB OSC-filled cavities display apparent lasing emission (also see Figure S7). The dissymmetry factor ($g$) for lasing is calculated as $g = 2 \times (I_L - I_R)/(I_L + I_R)$, where $I_L$ and $I_R$ correspond to the intensities of left- and right-handed polarization, respectively.[7b] The obtained $g$ for different cavity lengths are presented in Figure 4e. The $g$ value exhibits strong dependence on $\Delta\delta$, that is, $g$ decreases as $\Delta\delta$ increases. The maximum reached for $g$ in our system is as high as 1.2 at $\Delta\delta = 0$.

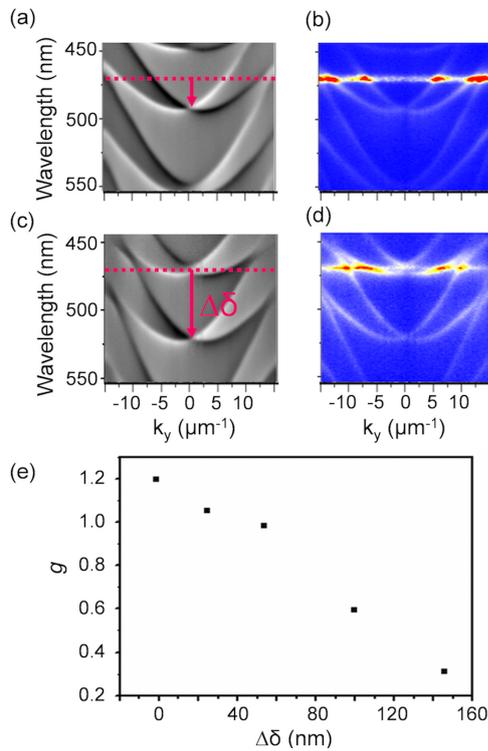

**Figure 4.** The ARR (a,c) and ARPL (b,d) spectra of the cavity with different lengths of 1085 nm and 1150 nm, respectively. (e) Dependence of the dissymmetry factor $g$ for lasing on the wavelength difference $\Delta\delta$.

## Conclusion

In summary, we have demonstrated CP lasing from pure organic crystal-filled microcavities without any chiral molecules or chiral structures. Based on the giant anisotropy and excellent lasing gain of the 6M-DSB single crystal, the tailored optical RD SOC gives rise to CP lasing in our OSC-filled microcavity. Experimental results indicate that the CP lasing emission with opposite helicities originates from the bottoms of the two spin-polarized valleys induced by RD SOC. Changing the cavity length can be used to tune the $g$ factor of the CP lasing which can reach a maximum value of 1.2 in our design. Our strategy breaks the limitation of inherent chiral conditions for the realization of CP lasers. This may provide a new idea for the design of CP lasers towards future 3D laser displays, information storage and other fields.

## Experimental Section

All other characterization data, original spectra, etc., have been provided in the Supporting Information.

## Acknowledgements

The authors thank Dr. H.W. Yin from ideaoptics Inc. for the support on the angle-resolved spectroscopy measurements.
This work was supported by the National Key R&D Program of China (Grant No. 2018YFA0704805, 2018YFA0704802 and 2017YFA0204503), the National Natural Science Foundation of China (22150005, 22090022, 21833005, 21873065 and 21790364), the Natural Science Foundation of Beijing, China (KZ202110028043), Beijing Talents Project (2019A23), Beijing Advanced Innovation Center for Imaging Theory and Technology.

**Keywords:** circularly polarized laser, organic crystals, optical microcavity, spin-orbit coupling

Supporting Information

# Circularly Polarized Lasing from a Microcavity Filled with Achiral Single-Crystalline Microribbons

Qian Liang,[a] Xuekai Ma,[b] Teng Long,[a] Jiannian Yao,[c] Qing Liao,[a,*] Hongbing Fu[a,*]

[a]Beijing Key Laboratory for Optical Materials and Photonic Devices, Department of Chemistry, Capital Normal University, Beijing 100048, People's Republic of China

[b]Department of Physics and Center for Optoelectronics and Photonics Paderborn (CeOPP), Universität Paderborn, Warburger Strasse 100, 33098 Paderborn, Germany

[c]Beijing National Laboratory for Molecular Sciences (BNLMS), Institute of chemistry, Chinese Academy of Sciences, Beijing 100190, People's Republic of China

E-mail: liaoqing@cnu.edu.cn; hbfu@cnu.edu.cn

## 1. Materials and Methods

### 1.1 Synthesis of 6M-DSB molecules

All starting materials were purchased from Sigma-Aldrich and used as received without further purification.

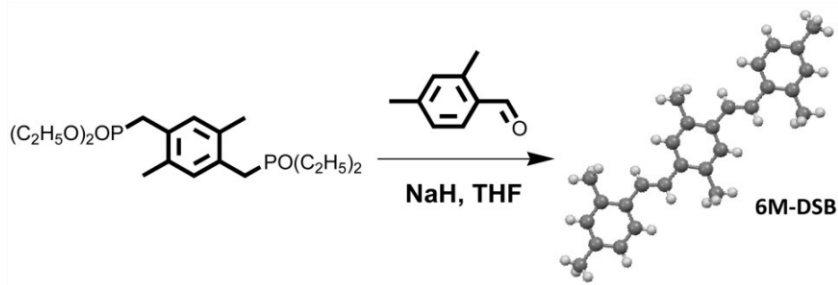

To a mixture of 2,5-dimethyl-1,4-xylene-bis(diethyl phosphonate) (0.81 g, 2.0 mmol) and the 2,4-dimethylbenzaldehyde (0.54 g, 4.0 mmol) in THF cooled in an ice bath was added 2 eq. NaH in small portions during a 30 min period. The reaction mixture was stirred at room temperature for 3 h and poured into water. The phase was extracted with $CH_2Cl_2$. The pooled organic phases were washed with water, dried over anhydrous $MgSO_4$, filtered, and evaporated. The product was separated by flash chromatography on silica gel by means of $CH_2Cl_2$/petroleum ether (1:4). Finally a highly fluorescent powder was obtained as the title compound (0.62 g, 1.7 mmol) in 85% yield. $^1H$ NMR (400 MHz, $CDCl_3$): δ7.51 (d, 2 H), 7.42 (s, 2 H), 7.18 (d, 4 H), 7.04 (d, 4 H), 2.43 (s, 6 H), 2.42 (s, 6 H), 2.34 (s, 6 H). ESI-HRMS: Calcd. For $[M+H]^+$: 367.5430. Found: 367.5376. Elemental Anal.: C, 91.82; H, 8.18.

### 1.2. Fabrication of single-crystal microribbons of 6M-DSB

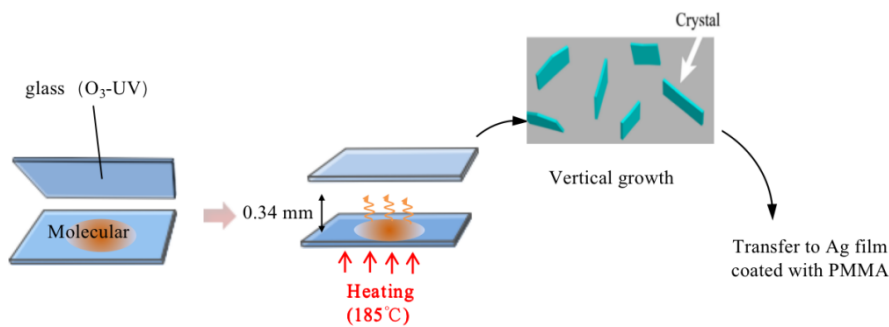

**Scheme S1. Schematic diagram of preparation of 6M-DSB microcrystals.**

The glass sheets (20mm×20mm) were successively cleaned with deionized water and anhydrous ethanol, respectively, and then were dried by $N_2$ and $O_2$ plasma for 10 min. The cleaned wafers were spincoated with 1 wt% solution of polystyrene in toluene. The thickness of the polystyrene film was 30 nm. The substrate was annealing in air atmosphere at 75 °C for more than one night, and then the 6M-DSB were laminated on it. The starting organic powders (typically much less than 0.5 mg) were put dispersedly. Then the substrates were placed on the hot stage, with a tiny space between the bottom and top ones separated by one small glass sleepers. The growth procedure was implemented by heating the bottom substrate to sublime the materials and formation of microcrystals on the down surface of the top substrate. The crucial parts of this method to obtain high quality single crystals are the spacing distance between the two substrates and setting suitable temperatures, which were found to be pivotal factors in determining the morphology of the grown crystals. The spacing distance of the two substrates were set at 340 μm in typical growth setup configuration of the text part, where micrometer-scale single crystals of 6M-DSB grew uniformly on the down surface of the top substrate under the condition of 185 °C. The single-crystal microribbons could be found on the surface of the $O_2$ plasma-modified of top substrate glass (F. Yin, *et al. New J. Chem.* **2020**, 44, 17552-17557.).

### 1.3. Microcavity fabrication

Firstly, we use the metal vacuum deposition system (Amostrom Engineering 03493) to thermally evaporate silver film with the thickness of 85 (±5) nm (reflectivity: R ≥ 99%) on the glass substrate, the root mean square roughness ($R_q$) of the silver film in the 5 μm×5 μm area is 2.45 nm, a 20 (±2) nm $SiO_2$ layer was deposited using vacuum Electron Beam evaporate on the silver film with $R_q$ of 2.31 nm, the deposited rate were both 0.2 Å/s and the base vacuum pressure is $3×10^{-6}$ Torr. Then the 6M-DSB crystals are transferred to a silver/$SiO_2$ film substrate, 20 (±2) nm $SiO_2$ and 35 (±2) nm (R ≈ 50%) silver was fabricated to form the microcavity. The 20 nm $SiO_2$ layers is used to prevent the fluorescence quenching of the 6M-DSB microbelt caused by directly contact of the metallic silver with the crystal.

### 1.4. Optical Spectroscopy Method

Note that in all spectral measurements in this work, the belt length was always projected to be perpendicular to the slit of the spectrometer.

The reflectivity measurement is shown in Scheme S2. The reflectivity was also measured using a Halogen lamp with the wavelength range of 400-700 nm. The light source was entered (cyan and white lines) and collected by using the same 100× microscope objective with a high numerical aperture (0.95 NA). The measurement angle can achieve ±71.8°. The yellow and white lines in Scheme S2 indicate the excited light path.

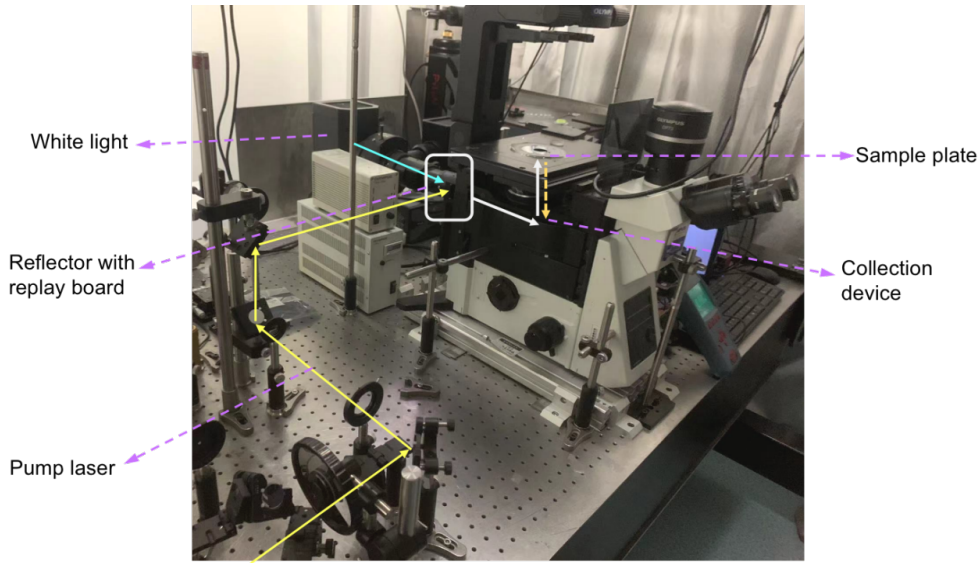

**Scheme S2.** The setup used in the experiment.

The reflectivity and photoluminescence spectroscopy were detected at room temperature in a home-made micro-area Fourier image which is presented to the spectrometer slit through four lenses, the schematic of the angle-resolved experimental setup is shown in Scheme S3. The reflected spectrum was collected by the spectrometer with a 300 lines/mm grating and a 400×1340 pixel liquid nitrogen cooled charge-coupled (CCD). For off-resonant optical pumping (400 nm, pulse width 150 fs) from a 1 kHz Ti: sapphire regenerative amplifier and 40-μm spot diameter with a near Gaussian beam profile, the angle-resolved photoluminescence shown was collected with the 300 lines/mm grating to improve the resolution of the spectrometer. In order to investigate the polarization properties, we placed a quarter-wave plate, a half-wave plate and a linear polarizer in the detection optical path in front of the linear polarizer to resolve the circular polarization in the left-handed and right-handed ($\sigma^+$ and $\sigma^-$) basis.

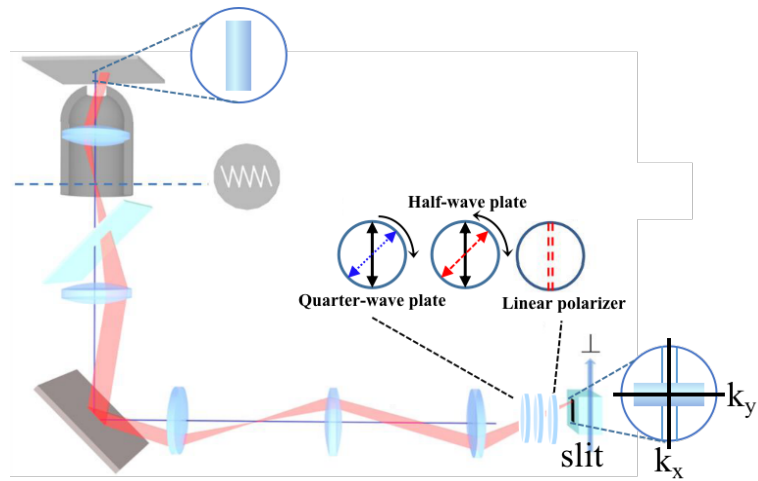

**Scheme S3.** The schematic diagram of the angle-resolved measurement setup.

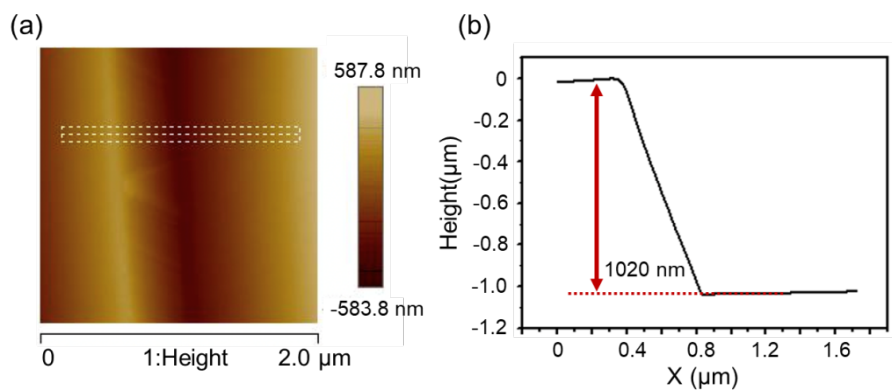

**Figure S1.** Atomic force microscope (AFM) of the 6M-DSB microcrystals. (a) AFM image and (b) the thickness of these microcrystals determined to be about 1 µm according to the AFM measurements.

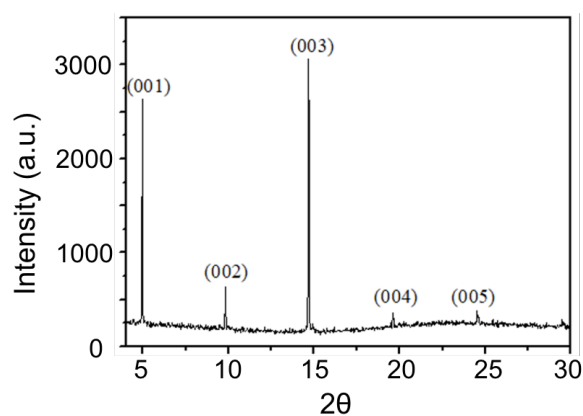

**Figure S2.** XRD spectrum of the microbelts showing a series of peaks corresponding to the (001) crystal plane with an interplanar spacing of 18.15 Å. The observation of high-order diffraction peaks, such as (002)-(005), suggests that the microribbons adopt a lamellar structure with the crystal (001) plane being parallel to the substrate (see Figure 1c).

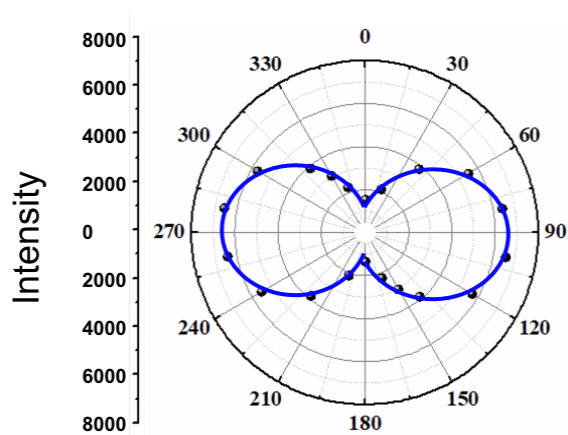

**Figure S3.** Photoluminescence polarization of the 6M-DSB crystal.

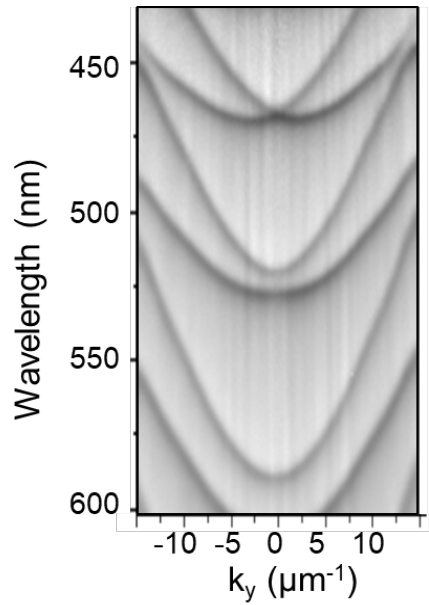

**Figure S4.** Unpolarized angle-resolved reflectivity of the optical microcavity filled by organic single crystals with the thickness of 1020 nm.

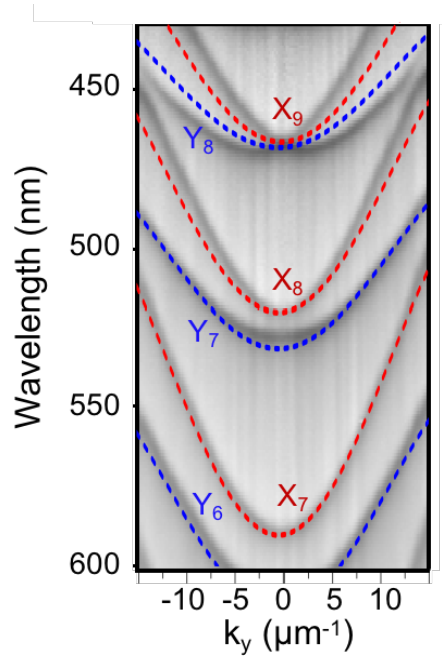

**Figure S5.** The corresponding refractive index of the two simulated cavity modes (Y polarized modes and X polarized modes). The refractive index corresponding to the red curve is 2.10 and to the blue curve is 2.50. Note that the middle two branches of the simulated results are slightly different from the corresponding ones shown in Figure 2d in the main text, which have been optimized by considering the RD effect, because in experiments, the RD effect contributes also to these two branches, although the influence is very weak.

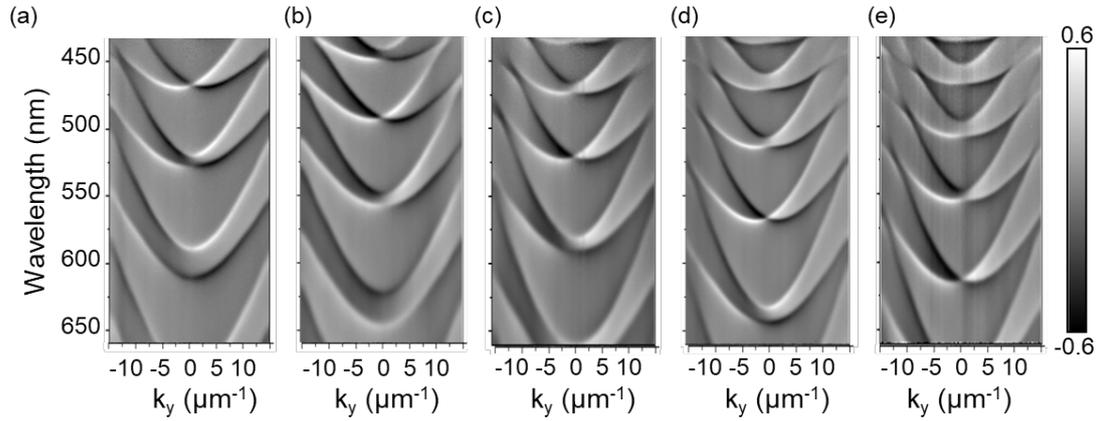

**Figure S6.** Angle-resolved reflectivity spectra at different cavity lengths: 1020 nm (a), 1085 nm (b), 1150 nm (c), 1260 nm (d), and 1330 nm (e).

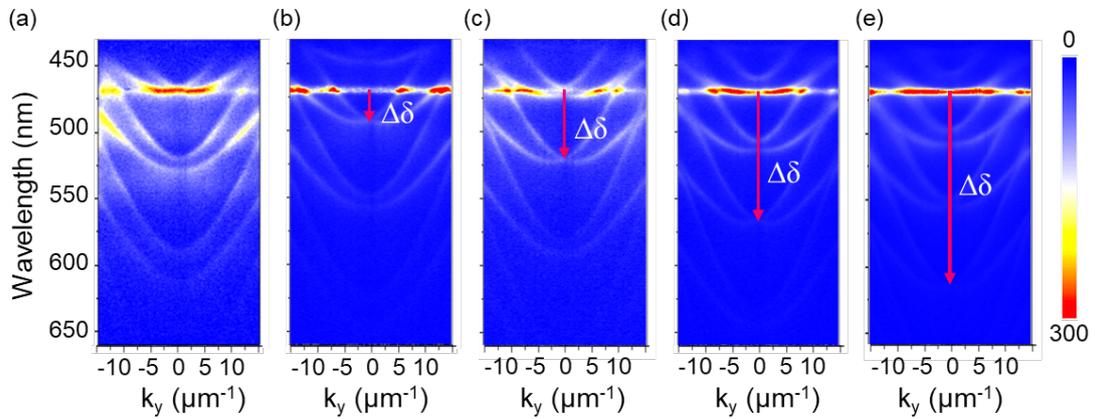

**Figure S7.** Angle-resolved PL spectra above the threshold at different cavity lengths, corresponding respectively to the reflectivity spectra shown in Figure S6. The corresponding $\Delta\delta$ is 0 nm (a), 24 nm (b), 53 nm (c), 99 nm (d), and 145 nm (e). It is clear that the *g* value increases with the decrease of $\Delta\delta$. The excitation powers for these samples are 3.032, 2.69, 2.04, 2.87, and 1.95 mJ/cm$^2$, respectively.